\documentclass{article}
\usepackage[margin=1in]{geometry}
\usepackage{graphicx}
\usepackage{amsmath}
\usepackage{amssymb}
\usepackage[numbers]{natbib}
\usepackage{authblk}

\usepackage{siunitx}                   
\usepackage{booktabs}                  
\usepackage{subcaption}                
\usepackage{placeins}  
\usepackage{xcolor}
\usepackage{adjustbox}
\usepackage{siunitx}
\usepackage{booktabs}
\usepackage{subcaption}
\usepackage{placeins}
\usepackage{tikz}
\usetikzlibrary{shapes.geometric, arrows.meta, positioning, calc, fit}
\usepackage{hyperref}
\usepackage{tikz}
\usetikzlibrary{shapes.geometric, arrows.meta, positioning, calc, fit}

\tikzstyle{block} = [rectangle, draw, fill=none,
    minimum width=9em, minimum height=3em, text centered, align=center]
\tikzstyle{circle_block} = [circle, draw, fill=none,
    minimum size=5.5em, text centered, align=center]
\tikzstyle{bias_block} = [circle, draw, fill=none,
    minimum size=5.5em, text centered, align=center]
\tikzstyle{line} = [draw, -{Latex[round]}]
\tikzstyle{surrogate_box} = [draw, dashed, inner sep=1em, rounded corners]
\title{Observation-Guided Neural Surrogate Learning for Scientific Simulation Emulation:\ A Single-Gauge Flood-Inundation Proof of Concept}
\author{Marzieh Alireza Mirhoseini\\
\texttt{malimirhoseini@gmail.com}}
\date{}

\begin{document}
\maketitle

\begin{abstract}
We present an observation-guided neural surrogate-learning framework for scientific simulation emulation, demonstrated on urban flood-inundation mapping. The framework combines LISFLOOD-FP hydrodynamic simulations with a real Gauge L stage record that is mapped to the simulation grid and converted to a datum-consistent local water-depth target before being used as single-site supervision. Focusing on a $256\times256$ crop around Gauge L in the Chicago metropolitan area, the method first constructs an ensemble-approximated Gaussian-process / local analogue surrogate (EnsCGP) to obtain a coarse flood-depth estimate and an uncertainty proxy. A U-Net--ASPP neural corrector then refines the coarse map using only simulation-derived and geospatial inputs: EnsCGP depth, the uncertainty proxy, rainfall, and spatial coordinates. The converted gauge-derived local depth is used only as a pointwise training target at the mapped gauge pixel; simulation-based losses are evaluated away from that pixel. Across temporally held-out events from 2013--2019, the emulator closely reproduces LISFLOOD-FP simulation targets outside the gauge-constrained pixel ($R^2\approx0.99$, mean absolute error $<0.01$ m) and shows strong pointwise consistency with the converted Gauge L local depth target under the stated rolling-year protocol. We interpret these results as strong simulator-emulation agreement with pointwise observation-guided correction, not as independent validation of real-world inundation accuracy or as a complete operational flood-forecasting system. The current study is intentionally limited to one gauge and one local region; broader validation with multiple gauges, independent flood-extent observations, ablation studies, wet-area/extent metrics, runtime benchmarking, and uncertainty calibration is required before operational claims.
\end{abstract}

\section{Introduction}
Flood risk assessment is critical for protecting lives, infrastructure, and economies against increasingly frequent and severe flooding events, exacerbated by climate change and rapid urban expansion \cite{Wing2017,Alfieri2017}. While flood risk assessment broadly focuses on understanding flood probabilities and potential impacts over extended periods, flood forecasting targets real-time or near-real-time predictions of imminent flood events. Accurate and timely flood forecasting remains challenging due to complex urban hydrodynamics, variable terrain characteristics, and uncertain meteorological inputs \cite{Hou2021,Zahura2022,Yan2021}.

High-fidelity hydrodynamic simulators such as LISFLOOD-FP and HEC-RAS solve shallow-water equations to simulate detailed flood dynamics, but their computational intensity limits real-time application and ensemble-based risk analysis, particularly for large urban domains \cite{Bates2010,Sharifian2023}. Comprehensive flood risk analyses often require many simulations across synthetic extreme-event scenarios, emphasizing the need for fast yet reliable surrogate models \cite{Bass2018}. In this preprint, the term ``rapid'' refers to replacing a new hydrodynamic solve with a trained surrogate evaluation. A hardware-specific emulator inference-time benchmark is not yet reported and should be added before stronger runtime claims are made.

Purely data-driven methods, including convolutional neural networks, ensemble learning, and random forests, offer rapid flood predictions, but they often suffer from poor generalization to unseen extreme events and can be weakly constrained at hydrometric gauge locations \cite{Liu2025,Hou2021,Zahura2022}. Hybrid surrogate modeling frameworks that combine physics-based simulations with machine learning have emerged as promising alternatives. For example, physics-guided surrogates integrating Gaussian-process models with coarse simulations have shown runtime reductions while maintaining accurate flood-extent and depth estimates relative to numerical targets \cite{Fraehr2024,Bass2018,Fraehr2022}. Physics-informed neural networks and neural-operator systems have also demonstrated fast hydrodynamic surrogacy in several settings \cite{Raissi2019,Donnelly2024,Xu2024}.

Despite these advances, current surrogate models rarely integrate ensemble-based uncertainty proxies, pointwise supervision from real gauge records, and targeted reduction of worst-case spatial errors within a single framework. These features are important for decision support because an otherwise accurate inundation map can still be operationally misleading if it is biased at a monitored gauge or if a small region exhibits a large local error. Capturing worst-case inundation behavior during extreme events remains difficult for standard training approaches, especially when historical simulation archives are limited. This manuscript extends our earlier few-data rainfall-driven inundation surrogate work for Chicago, which used EnsCGP-initialized deep learning and post-hoc bias correction, and which was also presented in preliminary form at AGU24 \cite{Mirhoseini2024FewData,MirhoseiniAGU2024}. The present work differs by using a U-Net--ASPP delta-correction architecture and by adding an observation-guided single-gauge training target.

In this study, we present observation-guided neural surrogate learning for scientific simulation emulation as a single-gauge, local proof of concept. The method combines two components: (1) an ensemble-approximated Gaussian-process / local analogue regression surrogate that provides a physically informed coarse inundation estimate, and (2) a deep convolutional neural network based on a U-Net architecture with an atrous spatial pyramid pooling (ASPP) module \cite{Ronneberger2015,Chen2018} that refines the coarse map while emphasizing large local errors through the composite objective described below. The real gauge observation is deliberately not supplied as a model input. Instead, the Gauge L stage record is first mapped to the corresponding simulation pixel using its grid coordinates and converted to a local water-depth target in the same units and vertical reference as the model. The loss function then uses this converted gauge-derived local depth only at the gauge pixel, while all other spatial errors are evaluated against the LISFLOOD-FP simulation target. Applied to historical flood events in the Chicago metropolitan region, the proposed EnsCGP--U-Net hybrid emulator shows strong agreement with the simulator outside the gauge-constrained pixel and improves pointwise consistency with the converted Gauge L local depth target under the stated rolling-year protocol. We therefore frame the contribution as observation-guided emulation of a hydrodynamic simulator, rather than independent validation of real-world flood depth or a fully validated operational flood-forecasting system.

\begin{figure*}[t]
  \centering
  \begin{subfigure}{0.49\textwidth}
    \includegraphics[width=\linewidth]{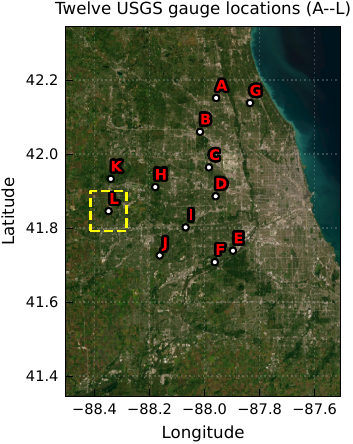}
    \caption{}
  \end{subfigure}
   \hfill
  \begin{subfigure}{0.49\textwidth}
    \includegraphics[width=\linewidth]{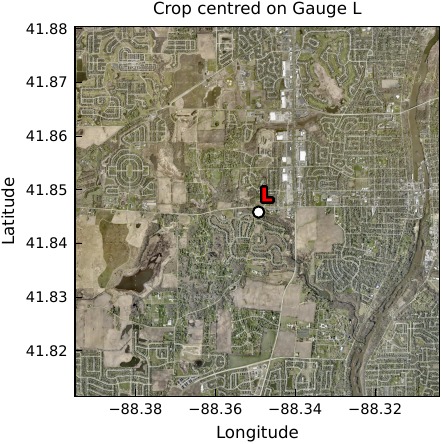}
    \caption{}
  \end{subfigure}  
  \caption{\textbf{Study area and gauge network.} (a) Twelve USGS river gauges (A--L) across the Chicago metropolitan region. (b) The $256\times256$ study crop centered on Gauge~L used for high-resolution emulation. Background satellite imagery is from Google Earth and is shown only for geographic context; no model inputs, training targets, or quantitative results are derived from this imagery. Google Earth and any embedded data-provider attribution should be retained as displayed in the source imagery. Coordinates are in degrees.}
   
\label{fig:region-gauge}
\end{figure*}
\section{Study area and data}
\subsection{Study area and gauge network}
The study domain is the greater Chicago metropolitan region in northeastern Illinois, USA, covering roughly 90 km north--south by 70 km east--west from the Lake Michigan shoreline to outlying rural catchments (Figure~\ref{fig:region-gauge}). Geographically, it spans approximately $41.345^\circ$--$42.345^\circ$N and $88.505^\circ$--$87.505^\circ$W. Twelve United States Geological Survey (USGS) hydrometric gauges, denoted A--L, monitor river stage across a range of urban and peri-urban sub-basins in this region. We simulated historical flood events from 1994--2019 using LISFLOOD-FP v8.1 \cite{Bates2010,Sharifian2023}, a GPU-accelerated finite-volume solver for the two-dimensional shallow-water equations. The model topography is defined by the approximately 30 m Shuttle Radar Topography Mission digital elevation model \cite{Farr2007}. Spatially distributed Manning roughness coefficients were assigned based on land-cover types from the 2016 National Land Cover Database \cite{Homer2020}. Meteorological forcing consisted of daily $1\,\mathrm{km}\times1\,\mathrm{km}$ gridded precipitation from Daymet v4 \cite{Thornton2022}.

For the gauge network, we retrieved observed daily water levels (river stage) from the National Water Information System using the hydrofunctions Python API \cite{USGSNWIS,Roberge2023}. Only the Gauge L record is used in the experiments below. The Gauge L coordinate is mapped to the corresponding pixel $p_g=(i_g,j_g)$ in the $256\times256$ simulation crop, and the observed stage record is converted to a local gauge-derived water-depth target in meters before it is used in training. This converted value is never included among the neural-network input channels. The exact USGS site identifier, parameter code, original units, datum metadata, grid-coordinate mapping, and preprocessing script should be released with the peer-reviewed version.

A USGS stage record is not automatically identical to gridded flood depth, so the Gauge L observation is made commensurate with the model output before it enters the loss. We first express the reported stage as an observed water-surface elevation in the model vertical reference frame and then subtract the local bed or ground elevation assigned to the mapped gauge pixel:
\begin{equation}
\eta_{\mathrm{obs}} = h_{\mathrm{stage}} + z_{\mathrm{datum}}, \qquad
 d_{\mathrm{obs}}(p_g)=\max\left(0,\eta_{\mathrm{obs}}-z_{\mathrm{local}}(p_g)\right),
\label{eq:stage_depth}
\end{equation}
where $h_{\mathrm{stage}}$ is the reported river stage, $z_{\mathrm{datum}}$ is the gauge-datum elevation transformed to the same vertical reference frame as the model, $z_{\mathrm{local}}(p_g)$ is the local bed or ground elevation assigned to the mapped gauge pixel, $\eta_{\mathrm{obs}}$ is the observed water-surface elevation, and $d_{\mathrm{obs}}(p_g)$ is the corresponding local observed water depth. The gauge term in the training objective uses $z_{\mathrm{obs}}(p_g)=d_{\mathrm{obs}}(p_g)$, so the comparison at $p_g$ is in meters of local depth, consistent with the model flood-depth output. Spatial-map accuracy away from $p_g$ is still evaluated as agreement with the LISFLOOD-FP simulation target; the gauge pixel is reported separately because its target comes from an observation rather than from the simulator. A later peer-reviewed version should report the full stage-to-depth conversion, vertical-datum handling, gauge-to-grid mapping, and conversion uncertainty.

\begin{figure*}[t]
  \centering
  \begin{subfigure}{0.49\textwidth}
    \includegraphics[width=\linewidth]{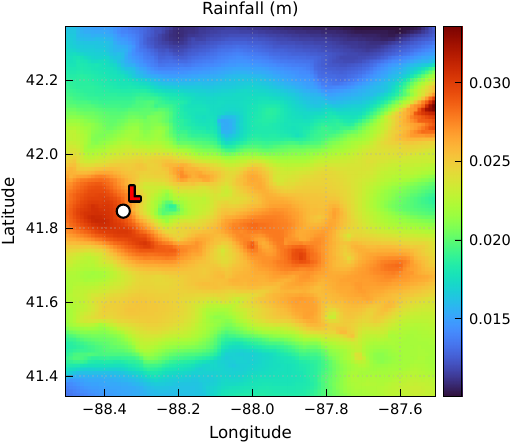}
    \caption{}
  \end{subfigure}
  \hfill
  \begin{subfigure}{0.49\textwidth}
    \includegraphics[width=\linewidth]{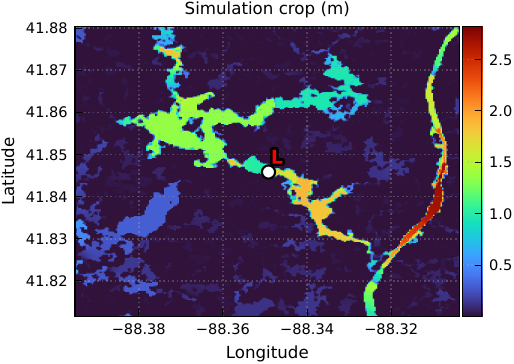}
    \caption{}
  \end{subfigure}
  \caption{\textbf{ Maximum daily rainfall in 2019 and simulated flood inundation (27 April 2019).} (a) Daily rainfall field on the wettest day of 2019. (b) Corresponding LISFLOOD-FP flood depth over the $256\times256$ crop; depths in meters. Gauge L location is marked in white circle.}
\label{fig:rainfall_flood}
\end{figure*}

All simulations and observations were aligned temporally so that each event corresponds to a day of intense rainfall and the associated simulated daily peak inundation response used as the learning target. This event-scale target implicitly includes routing and storage effects in the numerical simulation, but the neural-network input described below uses the event rainfall field and does not explicitly include antecedent soil moisture, drainage state, sub-daily rainfall intensity, lead-time uncertainty, or sensor latency. The current experiment should therefore not be interpreted as an operational nowcasting system. Figure~\ref{fig:rainfall_flood} provides a representative example of the event-scale input and target used in this study. It shows the daily rainfall field for the wettest 2019 event and the corresponding LISFLOOD-FP simulated flood-depth response over the $256\times256$ Gauge L-centered crop, illustrating how the rainfall forcing and simulation target are paired before being used in the surrogate-learning workflow.

\subsection{Focus region around Gauge L}
Figure~\ref{fig:region-gauge} illustrates the regional gauge network and the focused sub-domain around Gauge L, an urban tributary basin selected for high-resolution emulation. We cropped the full Chicago domain to a $256\times256$ grid around Gauge L for training the emulator. This sub-domain captures heterogeneous land cover and urban features while reducing the output dimensionality for the machine-learning model. By focusing on Gauge L, we can use an actual gauge-derived measurement from that site as a single-site supervision signal at the gauge pixel and evaluate how well the model reproduces surrounding flood patterns relative to the simulator without receiving the measured gauge value as an input. The present experiments are limited to this single local region and should be read as a proof of concept; extension to additional gauges, basins, and independent flood observations is necessary before making broad generalization claims.

The coordinate channels used below intentionally specialize the emulator to this fixed spatial crop. Transfer to another region would require retraining, domain adaptation, or validation demonstrating that the learned coordinate-dependent corrections generalize.
\begin{figure}[t]
    \centering
    \begin{adjustbox}{max width=\textwidth}
    \begin{tikzpicture}[
        font=\small,
        node distance=1.25cm and 1.25cm,
        >={Latex[length=2.2mm,width=1.6mm]},
        data/.style={
            rectangle, rounded corners=3pt, draw=black!60, fill=gray!8,
            line width=0.7pt, text width=2.55cm, minimum height=1.05cm,
            align=center, inner sep=5pt
        },
        model/.style={
            rectangle, rounded corners=3pt, draw=blue!65!black, fill=blue!6,
            line width=0.8pt, text width=2.95cm, minimum height=1.12cm,
            align=center, inner sep=5pt
        },
        obs/.style={
            rectangle, rounded corners=3pt, draw=green!50!black, fill=green!6,
            line width=0.75pt, dashed, text width=2.85cm, minimum height=1.05cm,
            align=center, inner sep=5pt
        },
        target/.style={
            rectangle, rounded corners=3pt, draw=blue!45!black, fill=blue!3,
            line width=0.75pt, dashed, text width=2.85cm, minimum height=1.05cm,
            align=center, inner sep=5pt
        },
        loss/.style={
            rectangle, rounded corners=3pt, draw=orange!85!black, fill=orange!10,
            line width=0.85pt, text width=3.05cm, minimum height=1.10cm,
            align=center, inner sep=5pt
        },
        flow/.style={->, line width=0.85pt, draw=black!75},
        train/.style={->, line width=0.85pt, dashed, draw=black!55},
        lab/.style={font=\scriptsize, align=center, text=black!70}
    ]

        \node[data]  (rain) {Gridded rainfall\\fields\\\textit{input}};
        \node[model, right=of rain] (enscgp) {EnsCGP local\\analogue surrogate\\\textit{coarse depth + uncertainty}};
        \node[model, right=of enscgp] (cnn) {U-Net--ASPP\\neural corrector\\\textit{depth, uncertainty, rainfall, coords}};
        \node[data, right=of cnn] (flood) {High-resolution\\flood-depth map\\\textit{output}};

        \draw[flow] (rain) -- (enscgp);
        \draw[flow] (enscgp) -- node[above, lab] {$z_{\mathrm{GP}}$ + uncertainty} (cnn);
        \draw[flow] (rain.south) to[out=-35,in=-145] node[below, lab] {rainfall channel} (cnn.south);
        \draw[flow] (cnn) -- (flood);

        \node[obs, below=1.45cm of rain] (gauge) {Gauge L stage\\record};
        \node[obs, below=1.45cm of enscgp] (convert) {Map to grid\\convert stage to\\local depth};
        \node[target, below=1.45cm of cnn] (sim) {LISFLOOD-FP\\target map\\\textit{non-gauge pixels}};
        \node[loss, below=1.45cm of flood] (lossnode) {Composite training loss\\gauge term at $p_g$\\spatial terms elsewhere};

        \draw[train] (gauge) -- node[above, lab] {datum consistent} (convert);
        \draw[train] (convert) -- (lossnode);
        \draw[train] (sim) -- (lossnode);
        \draw[train] (flood) -- (lossnode);
        \draw[train] (lossnode.north west) to[out=155,in=-55] node[above, lab] {training only} (cnn.south east);

        \node[draw=red!65!black, fill=red!3, rounded corners=3pt,
              font=\scriptsize, align=center, text width=8.8cm]
              at ($(gauge.south)!0.5!(sim.south)+(0,-0.75)$)
              {The real gauge value is \textbf{not} supplied as a neural-network input; it is used only as a pointwise training target at the mapped gauge pixel.};

    \end{tikzpicture}
    \end{adjustbox}
    \caption{\textbf{Two-stage observation-guided surrogate-emulation pipeline.}
    Gridded rainfall is first mapped to a coarse flood-depth estimate and an uncertainty proxy by the EnsCGP local analogue surrogate. The U-Net--ASPP corrector refines this estimate using the coarse depth, uncertainty proxy, rainfall field, and coordinate channels. During training only, the Gauge~L stage record is mapped to the simulation grid and converted to a datum-consistent local depth target. The composite loss uses this converted target at the mapped gauge pixel and LISFLOOD-FP simulation targets away from that pixel; the gauge value is not supplied as an input feature.}
    \label{fig:pipeline}
\end{figure}
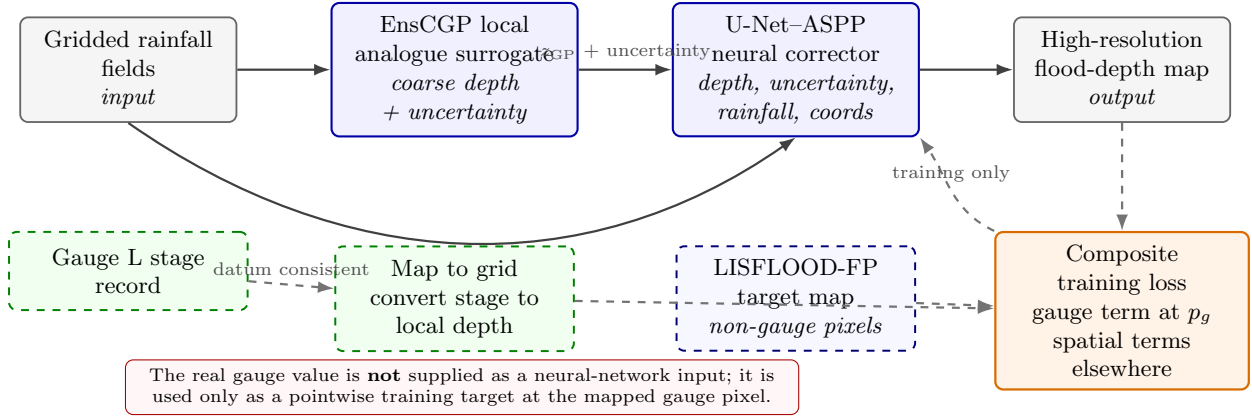
\section{Hybrid surrogate modeling approach}
As illustrated in Figure~\ref{fig:pipeline}, our approach consists of two stages. First, an ensemble-approximated Gaussian-process / local analogue surrogate takes gridded rainfall fields as input and produces a coarse initial flood-inundation map. Second, a deep convolutional neural network refines this initial guess to yield the final high-resolution flood-depth map. The real gauge observation is used only in the training loss: after stage-to-depth conversion, the mapped gauge pixel is compared with the converted gauge-derived local depth target rather than with the simulated value. The real gauge value is not passed to the CNN as an input feature.

\subsection{Ensemble-approximated Gaussian-process / local analogue surrogate}
Conventional Gaussian-process regression provides a principled way to predict a flood response $y_q$ for a new rainfall input $x_q$ by leveraging covariance structure in the training data. Let $X$ and $Y$ denote matrices of training inputs and outputs, respectively, where each column is one event. A classical linear conditional-mean predictor can be written schematically as
\begin{equation}
 y_q-\bar{Y} = \operatorname{Cov}(Y,X)\operatorname{Cov}(X,X)^{-1}(x_q-\bar{X}).
\label{eq:gp_conditional}
\end{equation}
Directly computing and inverting covariance matrices is infeasible for high-dimensional outputs. We therefore adopt a reduced-rank, ensemble covariance-regression approximation related to Gaussian-process flood surrogates, EOF/GP modeling, and fast ensemble-smoothing/data-assimilation ideas \cite{Fraehr2022,Fraehr2024,Ravela2007,Ravela2010}:
\begin{equation}
\operatorname{Cov}(X,X)\approx \frac{1}{N-1}(X-\bar{X})(X-\bar{X})^T, \qquad
\operatorname{Cov}(Y,X)\approx \frac{1}{N-1}(Y-\bar{Y})(X-\bar{X})^T .
\label{eq:cov_approx}
\end{equation}
In this implementation, this stage is best interpreted as an ensemble-approximated GP / local analogue regression surrogate rather than a fully calibrated probabilistic GP posterior or a full data-assimilation smoother. The ensemble-smoothing and real-time geophysical-estimation references provide methodological background for reduced-rank ensemble covariance estimation; the present flood emulator does not implement those algorithms exactly.

Rather than training a single global emulator, a new local model is built on the fly for each query event using only a small set of analogous past events. For a new rainfall event $x^{(q)}$, we flatten each training rainfall field into a vector and measure Euclidean distance to all training inputs. We select up to $K=100$ nearest events, subject to a distance threshold defined as the 90th percentile of minimum-neighbor distances in the training set. If fewer than three events satisfy the threshold, the three nearest neighbors are used. This procedure yields $n\leq K$ analogue pairs $\{x^{(i)},y^{(i)}\}_{i=1}^{n}$.

To make the dimensions explicit, let $d_x$ denote the number of rainfall-input features and $d_y$ the number of flood-output pixels. After centering the selected analogues, define
\begin{equation}
X'=[x^{(1)}-\bar{x},\ldots,x^{(n)}-\bar{x}]\in\mathbb{R}^{d_x\times n}, \qquad
Y'=[y^{(1)}-\bar{y},\ldots,y^{(n)}-\bar{y}]\in\mathbb{R}^{d_y\times n}.
\end{equation}
Within this local neighborhood, we approximate the relationship between rainfall anomalies and flood-response anomalies as linear, $Y'\approx A X'$, with $A\in\mathbb{R}^{d_y\times d_x}$. To regularize the least-squares fit, we compute a truncated singular value decomposition of the rainfall anomaly matrix,
\begin{equation}
X' = U\Sigma V^T,
\end{equation}
and retain the leading $r$ components that explain at least 99\% of the analogue-set energy. Denoting the truncated matrices by $U_r$, $\Sigma_r$, and $V_r$, the reduced linear mapping is
\begin{equation}
A = Y' V_r \Sigma_r^{-1} U_r^T,
\end{equation}
and the coarse prediction for the query event is
\begin{equation}
\hat{y}^{(q)} = \bar{y} + A\left(x^{(q)}-\bar{x}\right).
\end{equation}
The reported uncertainty channel is an analogue-ensemble standard-deviation proxy rather than a calibrated GP posterior variance. We therefore use it only as an informative input feature for the neural corrector and leave formal probabilistic calibration to future work.

\begin{figure}[t]
    \centering
    \begin{tikzpicture}[x=0.8cm, y=0.8cm, >=Latex]
        \tikzstyle{enc}=[draw, fill=blue!20, minimum width=1.8cm, minimum height=1.2cm, font=\footnotesize, align=center]
        \tikzstyle{dec}=[draw, fill=blue!20, minimum width=1.8cm, minimum height=1.2cm, font=\footnotesize, align=center]
        \tikzstyle{aspp}=[draw, fill=orange!20, minimum width=1.6cm, minimum height=0.8cm, font=\footnotesize, align=center]
        \tikzstyle{input}=[draw, fill=gray!30, minimum width=1.4cm, minimum height=0.5cm, font=\scriptsize, align=center]
        \tikzstyle{output}=[draw, fill=gray!30, minimum width=1.8cm, minimum height=0.6cm, font=\footnotesize, align=center]

        \node[enc] (enc0) at (0, 0) {Conv Block\\(256)};
        \node[enc] (enc1) at (0, -3) {Conv Block\\(512)};
        \node[enc] (enc2) at (0, -6) {Conv Block\\(1024)};
        \node[enc] (enc3) at (0, -9) {Conv Block\\(2048)};

        \node[aspp] (aspp) at (3, -11) {ASPP};

        \node[dec] (dec3) at (6, -9) {Up+Conv\\(1024)};
        \node[dec] (dec2) at (6, -6) {Up+Conv\\(512)};
        \node[dec] (dec1) at (6, -3) {Up+Conv\\(256)};
        \node[dec] (dec0) at (6, 0) {Up+Conv\\(256)};

        \node[draw, fill=blue!20, minimum width=0.9cm, minimum height=0.7cm, font=\scriptsize, align=center] (finalconv) at (8, 0) {1$\times$1\\Conv};
        \node[output, right=0.8cm of finalconv] (output) {Flood Depth Map};

        \node[input] (in_coarse) at (-3,  1.6) {EnsCGP depth};
        \node[input] (in_uncert) at (-3,  0.8) {Uncertainty};
        \node[input] (in_rain)   at (-3,  0.0) {Rainfall};
        \node[input] (in_x)      at (-3, -0.8) {$x$-coord};
        \node[input] (in_y)      at (-3, -1.6) {$y$-coord};

        \draw[->] (in_coarse) -- (enc0);
        \draw[->] (in_uncert) -- (enc0);
        \draw[->] (in_rain) -- (enc0);
        \draw[->] (in_x) -- (enc0);
        \draw[->] (in_y) -- (enc0);

        \draw[->] (enc0.south) -- node[right, font=\scriptsize]{MaxPool $2\times$} (enc1.north);
        \draw[->] (enc1.south) -- node[right, font=\scriptsize]{MaxPool $2\times$} (enc2.north);
        \draw[->] (enc2.south) -- node[right, font=\scriptsize]{MaxPool $2\times$} (enc3.north);
        \draw[->] (enc3.south) |- node[below right, font=\scriptsize]{MaxPool $2\times$} (aspp.west);

        \draw[->] (aspp.east) -| node[above, font=\scriptsize]{Up $2\times$} (dec3.south);

        \draw[->] (dec3.north) -- node[right, font=\scriptsize]{Up $2\times$} (dec2.south);
        \draw[->] (dec2.north) -- node[right, font=\scriptsize]{Up $2\times$} (dec1.south);
        \draw[->] (dec1.north) -- node[right, font=\scriptsize]{Up $2\times$} (dec0.south);

        \draw[->, dashed] (enc0.east) -- (dec0.west);
        \draw[->, dashed] (enc1.east) -- (dec1.west);
        \draw[->, dashed] (enc2.east) -- (dec2.west);
        \draw[->, dashed] (enc3.east) -- (dec3.west);

        \draw[->] (dec0.east) -- (finalconv.west);
        \draw[->] (finalconv.east) -- (output.west);

    \end{tikzpicture}
    \caption{\textbf{Schematic of the refinement CNN architecture (U-Net+ASPP).}
    The model takes a five-channel input---the coarse EnsCGP flood-depth prediction, an analogue-ensemble uncertainty proxy, the rainfall field, and spatial $x$- and $y$-coordinate channels---and produces a high-resolution flood-depth map. The U-Net encoder (left) applies convolutional blocks and $2\times$ pooling to downsample, while the decoder (right) uses upsampling and convolution to restore resolution. An ASPP module at the bottleneck (orange) captures multi-scale context. Dashed arrows indicate skip connections between encoder and decoder levels, preserving spatial detail. A final $1\times1$ convolution outputs the refined flood-depth prediction at each grid cell. The real gauge value is not provided as an input to the network; it is used only as a pointwise training target at the mapped gauge pixel.}
    \label{fig:refine_arch}
\end{figure}
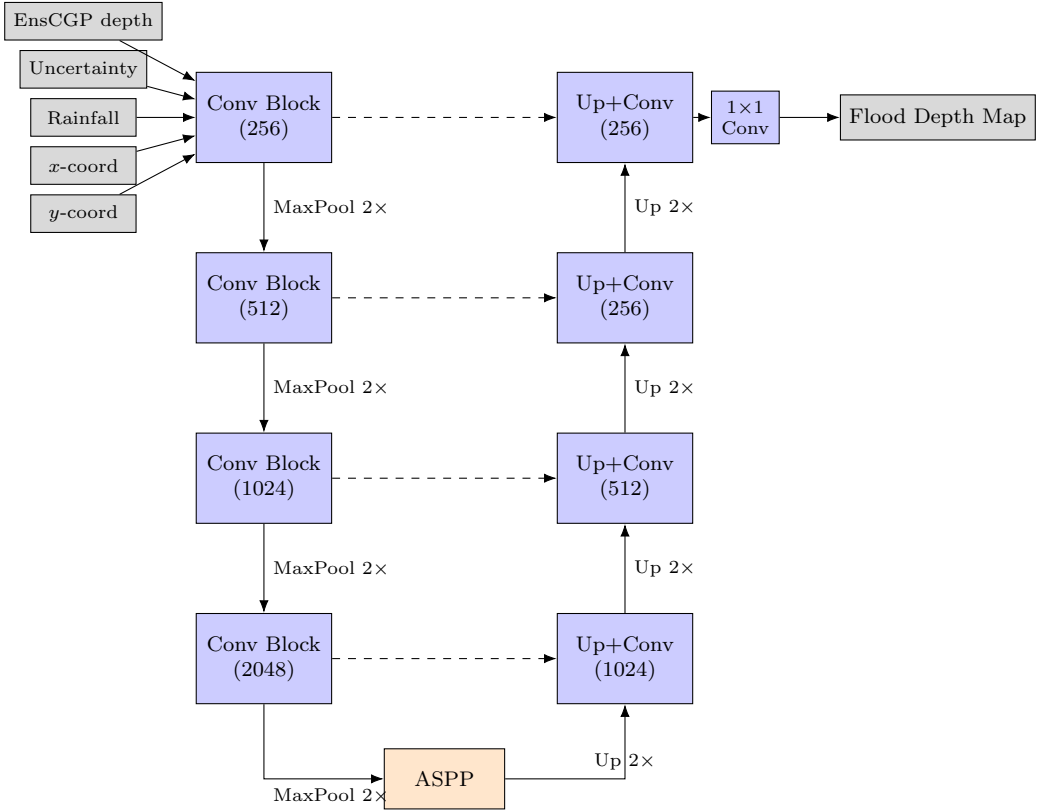
\section{Refinement CNN architecture and training}
\subsection{CNN input and architecture}
To refine the coarse maps produced by the EnsCGP surrogate, we employ a deep convolutional neural network with a U-Net backbone. All inputs are represented on the same $256\times256$ crop as the output. The model takes as input a five-channel tensor composed of (1) the coarse flood-depth prediction from EnsCGP, (2) the corresponding analogue-ensemble uncertainty proxy, (3) the daily rainfall field cropped and resampled to the output grid, and (4--5) spatial $x$ and $y$ coordinate channels (Figure~\ref{fig:refine_arch}). The coordinate channels are CoordConv-style features that intentionally give the convolutional network access to fixed spatial location information \cite{Liu2018CoordConv}; they are useful in a fixed crop but reduce translation invariance and therefore should not be interpreted as evidence of direct transferability to another basin. The real gauge value, the LISFLOOD-FP target map, and any held-out-year gauge information are excluded from the input tensor. Instead, the gauge observation enters only as a pointwise target in the loss function.

The CNN outputs a single-channel non-negative field with the same spatial dimensions as the input. Away from the gauge pixel, this output is interpreted as an emulator of LISFLOOD-FP flood depth in meters. At the gauge pixel, the output is compared with $z_{\mathrm{obs}}(p_g)=d_{\mathrm{obs}}(p_g)$, the Gauge L stage record converted to a local water-depth target in the same units and vertical reference as the model. Thus, the network output has a consistent depth interpretation over the full crop, but evaluation still reports the gauge pixel separately because its target is observational while the remaining pixels are supervised by LISFLOOD-FP.

The U-Net encoder applies successive convolutional blocks and $2\times$ downsampling to capture higher-level features, while the decoder uses transpose convolutions to restore the original resolution. An ASPP module at the bottleneck captures multi-scale context, which is useful for representing both broad flood extents and local details. Lateral skip connections concatenate encoder feature maps with decoder feature maps at corresponding resolutions. A final $1\times1$ convolution produces the refined flood-depth prediction at each grid cell.

\subsection{Gated delta correction mechanism}
Rather than predicting flood depth outright, the CNN learns to add a correction field over the coarse map. Let $z_{\mathrm{GP}}(i,j)$ denote the coarse EnsCGP prediction at pixel $(i,j)$. The CNN produces a correction $\Delta z(i,j)$ at the same location, and we combine these through a learned gate and a soft candidate-inundation mask. We introduce a trainable scalar parameter $\gamma$ and define $g=\sigma(\gamma)$, where $\sigma$ is the sigmoid function. The pre-activation refined field is
\begin{equation}
s(i,j)=z_{\mathrm{GP}}(i,j)+g\,M(i,j)\,\Delta z(i,j),
\end{equation}
and the final non-negative flood depth is
\begin{equation}
z_{\mathrm{pred}}(i,j)=\phi(s(i,j)), \qquad \phi(s)=\max(0,s).
\end{equation}
This zero-preserving non-negative activation is written explicitly because a standard Softplus would return a positive value at $s=0$ and could create artificial shallow water where the coarse prediction and correction are both zero.

Here $M(i,j)$ is a soft candidate-inundation mask derived deterministically from the coarse prediction to spatially modulate corrections. The mask is high where the EnsCGP predicts flooding and in a buffer around those regions, and decreases toward zero in far-field areas that are clearly dry according to the coarse surrogate. It is not interpreted as a hard physical constraint, and it does not use the real gauge value. Its purpose is to down-weight implausible far-field corrections while still allowing the network to repair missed floodplain pixels near the coarse inundation boundary or within low-lying connected areas. Because this mask affects where false negatives can be corrected, a reproducible version should report the wet-depth threshold, buffer or dilation radius, and smoothing rule used to construct $M$. False positives are controlled by the simulation-based loss terms and by evaluation against the LISFLOOD-FP target away from the gauge.

\subsection{Training objectives and curriculum}
We train the refinement CNN using a composite objective that balances spatial accuracy, worst-case performance, physically plausible corrections, and agreement with the converted gauge-derived local depth target at the monitored pixel. Let $z_{\mathrm{pred}}(p)$ denote the predicted flood depth at pixel $p$, $z_{\mathrm{sim}}(p)$ the LISFLOOD-FP target, $z_{\mathrm{GP}}(p)$ the EnsCGP estimate, $p_g$ the grid cell corresponding to Gauge L, and $z_{\mathrm{obs}}(p_g)=d_{\mathrm{obs}}(p_g)$ the converted gauge-derived local water-depth target at that location. Simulation-based losses are evaluated on the spatial domain with the gauge pixel excluded, $\Omega\setminus\{p_g\}$, whereas the gauge term uses the converted real observation:
\begin{equation}
\mathcal{L}=\lambda_1\mathcal{L}_{\mathrm{softmax}}^{\Omega\setminus p_g}+\lambda_2\mathcal{L}_{\mathrm{no\mbox{-}harm}}^{\Omega\setminus p_g}+\lambda_3\mathcal{L}_{\mathrm{top\mbox{-}k}}^{\Omega\setminus p_g}+\lambda_4\mathcal{L}_{\mathrm{align}}^{\Omega\setminus p_g}+\lambda_5\mathcal{L}_{\mathrm{gauge}}.
\end{equation}
For clarity, let $e(p)=z_{\mathrm{pred}}(p)-z_{\mathrm{sim}}(p)$, $e_{\mathrm{GP}}(p)=z_{\mathrm{GP}}(p)-z_{\mathrm{sim}}(p)$, $c(p)=gM(p)\Delta z(p)$, and $c^*(p)=z_{\mathrm{sim}}(p)-z_{\mathrm{GP}}(p)$. The intended non-gauge loss definitions are
\begin{align}
\mathcal{L}_{\mathrm{softmax}}^{\Omega\setminus p_g} &= \tau \log\sum_{p\in\Omega\setminus\{p_g\}} \exp\left(\frac{|e(p)|}{\tau}\right),\\
\mathcal{L}_{\mathrm{no\mbox{-}harm}}^{\Omega\setminus p_g} &= \frac{1}{|\Omega|-1}\sum_{p\in\Omega\setminus\{p_g\}} \left[\max\left(0, |e(p)|-|e_{\mathrm{GP}}(p)|\right)\right]^2,\\
\mathcal{L}_{\mathrm{top\mbox{-}k}}^{\Omega\setminus p_g} &= \frac{1}{k}\sum_{p\in\operatorname{TopK}_{\Omega\setminus\{p_g\}}(|e|)} |e(p)|,\\
\mathcal{L}_{\mathrm{align}}^{\Omega\setminus p_g} &= \frac{1}{|\Omega|-1}\sum_{p\in\Omega\setminus\{p_g\}} \rho\left(c(p)-c^*(p)\right),
\end{align}
with $\tau$ controlling the soft-maximum sharpness and $k$ defining the number of largest-error non-gauge pixels. The gauge loss is
\begin{equation}
\mathcal{L}_{\mathrm{gauge}}=\rho\left(z_{\mathrm{pred}}(p_g)-z_{\mathrm{obs}}(p_g)\right),
\end{equation}
where $\rho(\cdot)$ is implemented as an absolute, squared, or robust Huber penalty. Thus, the gauge pixel is not trained against the simulated water depth; it is trained against the converted gauge-derived local depth obtained from the datum-consistent stage conversion in Equation~\eqref{eq:stage_depth}. Because the observation is not provided as an input channel, this term does not allow inference-time copying of the real gauge value; instead, it encourages the learned correction field to reduce systematic local simulator bias at the monitored location. The exact choices of $\tau$, $k$, $\lambda_i$, and the schedule for $\rho$ should be released with the source code for full reproducibility.

The relative weights in the composite objective are adjusted during training. We begin with a relatively soft worst-case term and gradually sharpen the focus on high-error pixels. We similarly ramp up false-positive penalties over time: early training favors recall of inundated areas, whereas later stages suppress spurious water and improve precision. Event-level weighting gives additional emphasis to large or extreme floods. During an initial warm-up, the delta gate is clamped or biased so that $g\approx0$, keeping the output close to the EnsCGP baseline while the network learns useful features. After warm-up, the gate constraint is relaxed so that the CNN can apply larger corrections.

\subsection{Data-leakage controls and evaluation protocol}
Because the gauge-consistency result is evaluated against a converted gauge-derived local depth target, the temporal split and information flow are central to the scientific interpretation. For each held-out test year, the model is trained only on events from earlier years. Gauge values from the held-out year are not supplied as network inputs and are not used to compute the training loss for that split. Analogue selection for the EnsCGP stage is restricted to the training archive available before the test year. All reported simulator-emulation spatial metrics exclude the gauge pixel because that location has an observational target rather than the LISFLOOD-FP simulated target.

For the present arXiv proof of concept, the intended leakage control is therefore temporal: future-year observations and future-year simulation targets are withheld from the corresponding rolling training split. Any preprocessing quantities that are learned from data, including channel normalization, target scaling, analogue-distance standardization, and loss-scaling constants, should be fitted within the training archive for each split and then applied unchanged to the held-out year. If any implementation uses all-year statistics, that step should be treated as a potential leakage source and corrected before peer-reviewed submission.

This protocol is intended to prevent direct copying of test gauge values, but it does not remove all possible sources of optimism. Adjacent days from the same storm, preprocessing choices, event-selection thresholds, and hyperparameter tuning can still affect apparent generalization. Events from the same multi-day storm may introduce temporal dependence; storm-grouped splits are therefore needed before making stronger generalization claims. The present results should therefore be interpreted as a rolling-year proof of concept rather than as a fully leakage-audited operational validation. A peer-reviewed version should include a full event list, storm-grouped splits, normalization constants, random seeds, and negative-control tests such as shuffled gauge targets or a no-gauge-loss ablation.

\subsection{Current ablation and validation status}
The present arXiv version reports the main proof-of-concept experiment: LISFLOOD-FP simulation emulation outside the gauge pixel and pointwise gauge-guided correction at Gauge L. It does not yet include a complete ablation suite or independent spatial validation against satellite, aerial, or field-observed inundation products. Accordingly, the strong single-gauge consistency result is treated as hypothesis-generating rather than as definitive proof of observation skill. The minimum controls needed before stronger claims are: no-gauge-loss training, shuffled-gauge-target training, EnsCGP-only and rainfall-only baselines, U-Net without ASPP, direct-depth prediction versus delta correction, and removal of coordinate and uncertainty channels. Additional validation should include leave-one-gauge-out testing and independent comparison with remotely sensed or field-observed flood extent.

\begin{figure*}[t]
    \centering
    \begin{subfigure}{0.4\textwidth}
        \includegraphics[width=\linewidth]{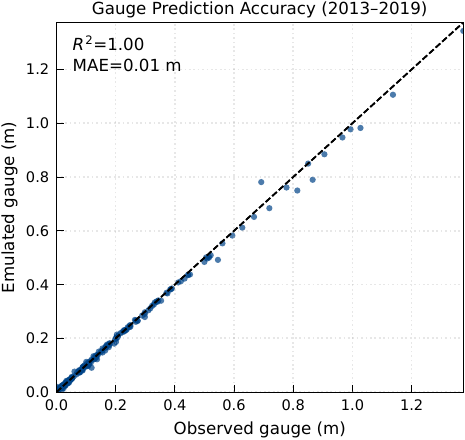}
        \caption{}
    \end{subfigure}
    \begin{subfigure}{0.58\textwidth}
        \includegraphics[width=\linewidth]{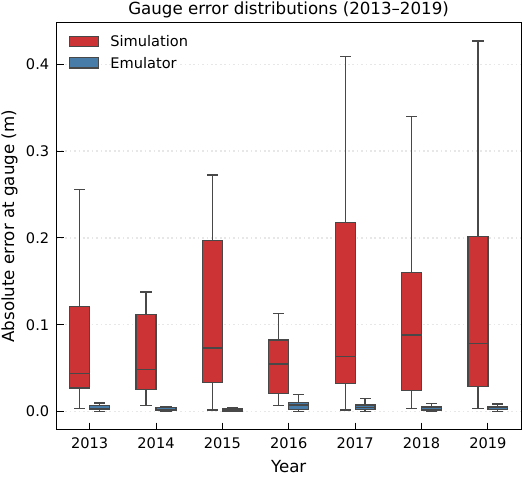}
        \caption{}
    \end{subfigure}\hfill
    \caption{\textbf{Gauge-derived depth consistency.} (a) Emulator-predicted local depth at Gauge~L versus the converted gauge-derived local depth target for all test events (2013--2019). The dashed line denotes the 1:1 reference line; the rounded statistics are $R^2 = 1.00$ and $\mathrm{MAE} \approx 0.01$ m, indicating strong pointwise agreement under the stated rolling-year protocol. (b) Absolute prediction error at Gauge~L for each year (2013--2019), comparing the physics-based simulation (red) to the emulator (blue), both evaluated against the converted gauge-derived local depth target. The emulator's gauge errors are small for the events shown, whereas the simulation often shows larger local deviations, up to approximately $\sim$0.4 m.}
    \label{fig:gauge_dis}
\end{figure*}
\begin{figure}[t]
    \centering
    \includegraphics[width=\linewidth]{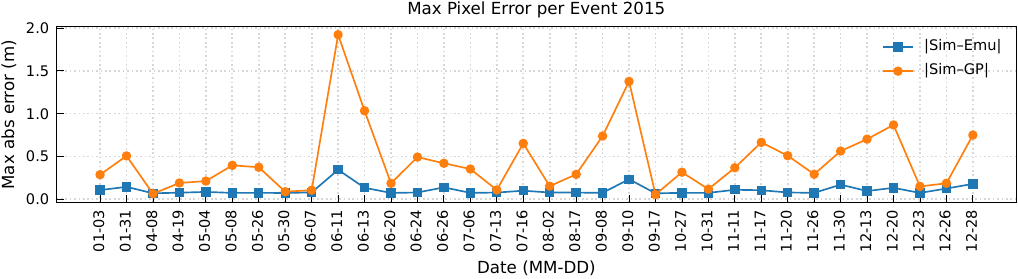}
    \caption{\textbf{Per-event maximum error for 2015 floods.} For each test event in 2015, the maximum absolute pixel error is plotted for the emulator (blue) and for the baseline coarse GP surrogate (EnsCGP, orange). Each point represents one flood event (date indicated along the horizontal axis). The gauge pixel is excluded from these error calculations. The emulator consistently yields much lower worst-case errors than the coarse surrogate across all events in 2015.}
    \label{err-2015}
    \end{figure}
    
\section{Results}
In our experiments, we trained the model on all events up to a given year and evaluated it on floods from the subsequent year, repeating this rolling evaluation for 2013 through 2019. This split simulates a scenario in which the emulator is periodically updated with newly available simulations and historical gauge records, then applied to future rainfall events without requiring the future real gauge value as an input. Overall, the hybrid model exhibits strong agreement with LISFLOOD-FP simulation targets across this held-out test set, while the gauge pixel is evaluated separately against the converted gauge-derived local depth target.

\begin{figure}[t]
    \centering
    \includegraphics[width=\linewidth]{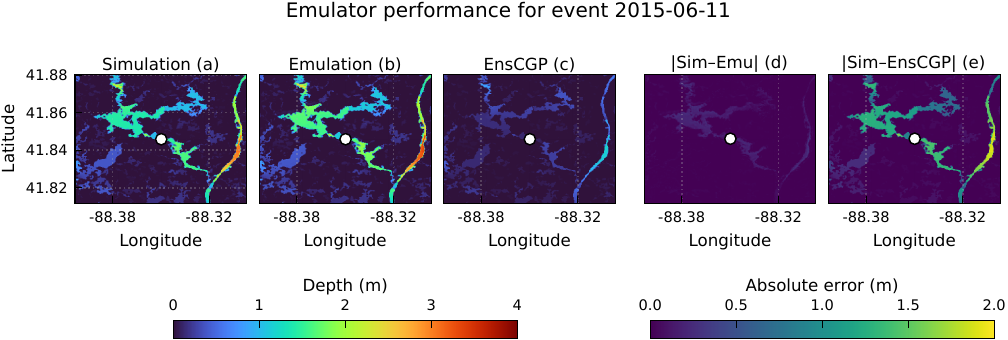}
    \caption{\textbf{Simulation, emulator, and initial-guess inundation comparison in an extreme 2015 flood.} Flood depth maps for the extreme event on 11~June~2015, comparing the LISFLOOD-FP simulation, the emulator, and the coarse EnsCGP surrogate (initial guess), with their respective error maps relative to the simulation. In this case, the EnsCGP surrogate severely \emph{underestimates} the flood extent (producing insufficient inundation), while the emulator substantially reduces these errors, yielding a flood map that more closely matches the LISFLOOD-FP simulation outside the gauge-constrained pixel.}
    \label{fig:max_error_pixelwise_2015}
\end{figure}

\begin{figure}[t]
    \centering
    \includegraphics[width=\linewidth]{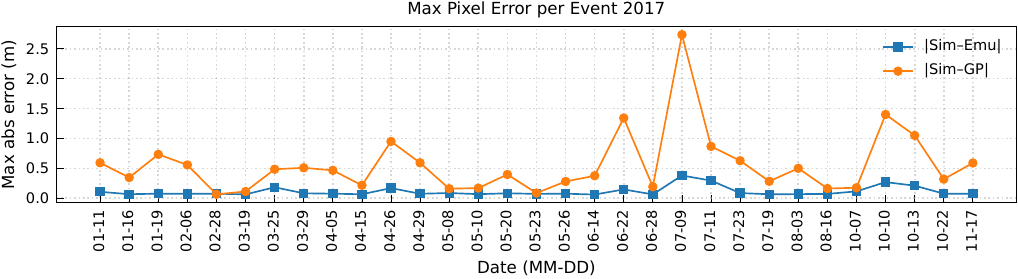}
    \caption{\textbf{Per-event maximum error for 2017 floods.} Similar to Figure~\ref{err-2015}, showing the maximum absolute pixel error for each test event in 2017 for the emulator (blue) and the EnsCGP surrogate (orange). The gauge pixel is excluded. The emulator's worst-case error is substantially lower than that of the coarse surrogate for every event in 2017.}
    \label{fig:err-2017}
    \end{figure}

     \begin{figure}[t]
     \centering
     \includegraphics[width=\linewidth]{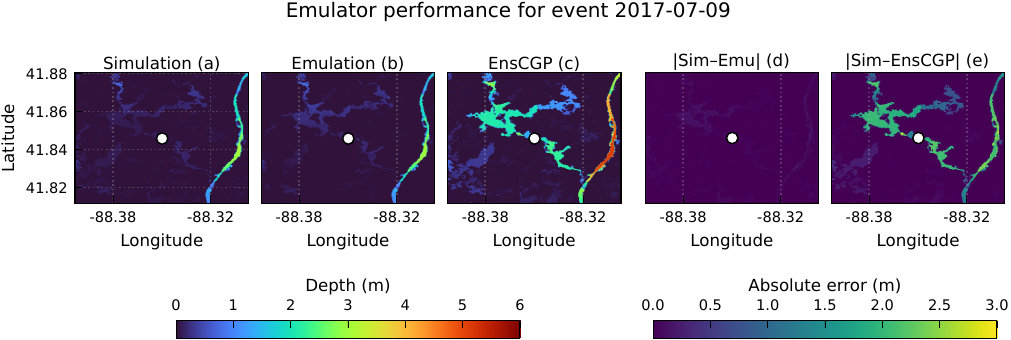}
     \caption{\textbf{Simulation, emulator, and initial-guess inundation comparison in an extreme 2017 flood}. Flood depth maps for the extreme event on 09 July 2017 from the LISFLOOD-FP simulation, the emulator, and the coarse EnsCGP surrogate, with their absolute error maps relative to the simulation. In this case, the EnsCGP surrogate substantially overestimates the flood extent, producing spurious inundation. The emulator successfully removes these errors, yielding a flood map that aligns closely with the high-fidelity simulation.}
     \label{fig:max_error_pixelwise_2017}
\end{figure}

\subsection{Gauge-derived depth consistency}
We first assess pointwise agreement at Gauge L, using the converted gauge-derived local depth $d_{\mathrm{obs}}(p_g)$ as the reference rather than the LISFLOOD-FP value at that pixel. The emulator predictions at Gauge L align closely with this single-site observational target, yielding a rounded $R^2=1.00$ and MAE $\approx0.01$ m under the rolling-year protocol (Figure~\ref{fig:gauge_dis}). In contrast, the LISFLOOD-FP simulation often exhibits larger local discrepancies relative to the same converted gauge-depth target. Because the real gauge value is not supplied as a network input, this agreement reflects observation-guided training at the gauge pixel rather than direct inference-time copying. Nevertheless, the very high gauge score should be interpreted cautiously: it demonstrates that the training objective can enforce strong single-site consistency under the stated protocol, not independent validation of the full spatial flood field. The result requires no-gauge-loss, shuffled-target, and storm-grouped controls before it can be interpreted as robust observational skill.

\subsection{Emulator refinement of the EnsCGP initial guess}
To evaluate how the emulator corrects the coarse analogue surrogate, we focus on worst-case errors and challenging case studies. For each held-out event we compute the per-event maximum absolute pixel error
\begin{equation}
e_{\max}=\max_{p\in\Omega\setminus\{p_g\}} |z_{\mathrm{sim}}(p)-\hat{y}(p)|,
\end{equation}
where $\hat{y}(p)$ is either the emulator or EnsCGP prediction. The gauge pixel is excluded because that location is supervised by the converted gauge-derived local depth target rather than by the simulated value. Figures~\ref{err-2015} and \ref{fig:err-2017} compare the per-event maximum absolute error of the emulator with that of the coarse EnsCGP surrogate for the 2015 and 2017 held-out test years. In both years, the emulator consistently reduces the largest non-gauge pixel errors relative to the EnsCGP baseline, indicating that the neural correction stage improves not only aggregate simulator agreement but also worst-case local behavior.

\begin{figure*}[t]
    \centering
    \begin{subfigure}[b]{0.48\textwidth}
        \centering
        \includegraphics[width=\linewidth]{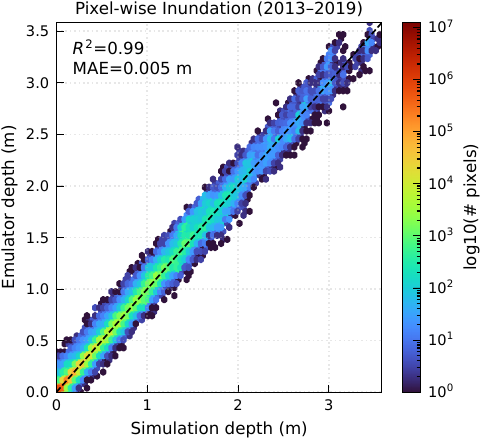}
        \caption{}
        \label{fig:scatter_sim_emu}
    \end{subfigure}
    \hfill
    \begin{subfigure}[b]{0.48\textwidth}
        \centering
        \includegraphics[width=\linewidth]{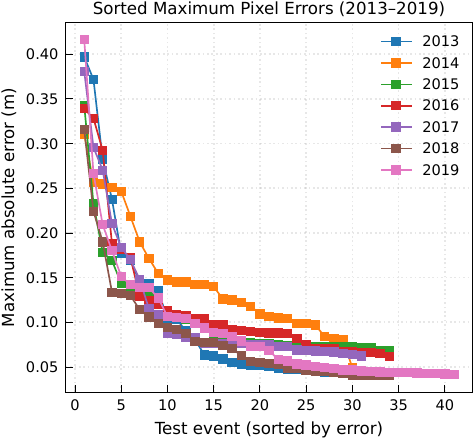}
        \caption{}
        \label{fig:sorted_max_pixel_error}
    \end{subfigure}
    \caption{\textbf{Emulator accuracy across held-out events (2013--2019).} (a) Pixel-wise comparison of emulator-predicted vs.\ simulated flood depths, shown as a density scatter plot (log$_{10}$ of pixel count). The dashed line indicates perfect agreement (1:1). The emulator achieves $R^2 \approx 0.99$ and overall $\mathrm{MAE} < 0.01$~m, indicating strong agreement with the simulator under the stated evaluation protocol. (b) Per-event maximum absolute error in flood depth for each test year (2013--2019), with events sorted in descending order of error. Over seven years of unseen floods, the emulator's worst-case error never exceeds $\sim$0.4~m. All statistics are computed after masking out the gauge pixel (which is directly constrained by observations).}
    \label{fig:emulator_eval}
\end{figure*}

The qualitative behavior matches these aggregates. For the 11 June 2015 extreme event (Figure~\ref{fig:max_error_pixelwise_2015}), the EnsCGP surrogate underestimates the inundation, whereas the emulator substantially reduces this bias and recovers a flood extent and depth field close to the LISFLOOD-FP simulation outside the observation-constrained gauge pixel. This recovery should be interpreted together with the soft-mask design above: the mask is not a hard binary restriction, so corrections can occur in buffered candidate-inundation areas rather than only in pixels already wet in the coarse surrogate. For the 2017 events summarized in Figure~\ref{fig:err-2017}, the EnsCGP surrogate tends to over-predict flood extent in the highest-error cases, producing spurious water that is substantially reduced by the emulator. A representative example of this behavior is shown in Figure~\ref{fig:max_error_pixelwise_2017} for the 09 July 2017 event. In this case, the EnsCGP initial guess produces broad spurious inundation, whereas the emulator removes much of this overprediction and produces a flood-depth field that more closely matches the LISFLOOD-FP simulation outside the gauge-constrained pixel.

\subsection{Overall simulator-emulation accuracy and extremes}
The hybrid emulator closely reproduces the LISFLOOD-FP simulation targets outside the gauge pixel. Across all held-out events ($n\approx240$; 2013--2019, with 32--42 events per year), pixel-wise predictions show strong agreement with simulations outside the gauge pixel (Figure~\ref{fig:emulator_eval}). The coefficient of determination is approximately $R^2\approx0.99$ and the overall mean absolute error is approximately 0.005 m. These aggregate pixel-wise metrics demonstrate simulator-emulation fidelity, but they can be dominated by dry or low-depth background cells in inundation problems. They should therefore not be interpreted as independent flood-inundation validation. Wet-pixel-only errors and flood-extent metrics such as precision, recall, F1/critical-success-index, and intersection-over-union at explicit wet/dry thresholds should be reported in a later validation-oriented version.

Beyond aggregate accuracy, we assess whether the model introduces large local errors under extreme conditions. For each event, we compute the maximum absolute error over all non-gauge pixels and sort events by this statistic (Figure~\ref{fig:emulator_eval}). Across seven years of held-out floods, the worst-case error never exceeds approximately 0.4 m. The distribution is also well behaved: the median per-event maximum error lies below 0.1 m for most years, indicating that at least half of the events have their largest non-gauge pixel discrepancy confined to a few centimeters. All spatial statistics exclude the gauge pixel, which is evaluated separately against the converted gauge-derived local depth target.

\FloatBarrier
\subsection{Novelty and implications}
Compared with many surrogate flood-mapping approaches, the distinctive feature of this proof of concept is the use of a real gauge-derived record as an external pointwise supervision signal while deliberately excluding the gauge value from the network input. Traditional data-driven flood models either omit point observations or treat them as input features, which can make gauge-site accuracy partly dependent on copying the supplied observation. Recent surrogate and graph-based flood-prediction approaches have made substantial progress in reducing overall error metrics \cite{Yang2024,Farahmand2023}. By contrast, our emulator combines a coarse simulation-informed surrogate, a neural correction field, worst-case pixel losses, and a gauge-observation loss that compares the predicted gauge pixel with a datum-consistent, gauge-derived local depth target rather than with the simulator. The result is a promising route toward pointwise observation-guided correction, but the present evidence remains limited to one gauge and simulator-based spatial targets away from that gauge.

\section{Discussion and limitations}
This arXiv version is intended to make the work visible and reproducible enough for technical discussion while clearly separating demonstrated results from future claims. The demonstrated result is that a neural corrector can closely emulate LISFLOOD-FP flood-depth maps in a Gauge L-centered crop and can incorporate a datum-consistent gauge-derived local depth target as pointwise supervision without receiving that gauge value as an inference-time input. The main limitation is that the spatial field away from the gauge is still supervised primarily by the simulator, so high agreement with LISFLOOD-FP should be interpreted as emulator fidelity rather than independent proof of real-world inundation accuracy.

\subsection*{Gauge target and physical interpretation} The gauge term is central to the proposed formulation, so the stage-to-depth conversion and gauge-to-grid mapping must be stated explicitly. A river-stage record is not automatically the same quantity as gridded flood depth; therefore, this manuscript maps the Gauge L coordinates to the simulation pixel $p_g$ and converts the USGS stage record to a local observed depth using Equation~\eqref{eq:stage_depth}. This gives the gauge loss the same unit and physical interpretation as the model flood-depth output. The gauge pixel is still reported separately from the non-gauge spatial metrics because its target is observational, whereas the rest of the map is supervised by LISFLOOD-FP. A later journal version should document the exact USGS site identifier, parameter code, unit conversion, vertical-datum transformation, local elevation choice, gauge-to-grid mapping, and uncertainty of the conversion.

\subsection*{Single-gauge and single-region scope} The present implementation uses a single monitored location, Gauge L, and a single $256\times256$ crop. This is useful for isolating the training design, but it does not establish general performance across gauges, catchments, or hydrologic regimes. The natural next step is multi-gauge supervision with leave-one-gauge-out validation, followed by testing in additional urban basins.

\subsection*{Simulator bias and independent validation} Because most pixels are trained against LISFLOOD-FP, the emulator can inherit simulator biases away from the gauge. Independent flood observations, such as satellite SAR inundation extent, aerial imagery, high-water marks, or vetted field reports, are required to evaluate real-world spatial accuracy. Until such validation is added, the strongest supported claim is rapid simulation emulation with a pointwise observational consistency constraint.

\subsection*{Spatial metrics and wet/dry thresholds} The present aggregate pixel-wise metrics should be read as simulator-emulation metrics. For flood-inundation validation, the wet/dry threshold used to define inundation should be reported explicitly, and metrics should be computed on wet pixels and flood extent as well as on all pixels. Useful follow-up diagnostics include wet-pixel MAE/RMSE, false-positive and false-negative area, precision, recall, F1/critical-success-index, and intersection-over-union at one or more depth thresholds.

\subsection*{Leakage, ablations, and robustness checks} The strong gauge score is encouraging but also demands careful leakage controls. The current protocol withholds future-year gauge values from training for each rolling split, but a peer-reviewed submission should include the exact event list, storm-grouped splits, normalization procedure, hyperparameter-selection protocol, and negative controls. Especially important are no-gauge-loss and shuffled-gauge-target controls, because they directly test whether the gauge term adds information and whether the single-site score could arise from leakage, temporal autocorrelation, or overfitting. Additional ablations should remove the uncertainty and coordinate channels, use a plain U-Net without ASPP, compare delta correction with direct-depth prediction, and benchmark against rainfall-only, EnsCGP-only, and standard machine-learning baselines.

\subsection*{Static geospatial inputs and urban hydraulics} The hydrodynamic simulations use a 30 m SRTM-derived elevation model and Manning roughness values assigned from the 2016 NLCD land-cover product. This setup is acceptable for a local proof of concept, but it cannot resolve all urban drainage, curb, culvert, bridge, building, and sewer-network controls on flood depth. It also uses a static 2016 land-cover representation for events spanning 1994--2019, which may introduce historical land-cover mismatch. These limitations affect the LISFLOOD-FP targets and therefore can be inherited by the emulator.

\subsection*{Forcing resolution and operational use} The experiments use daily Daymet precipitation and target daily peak simulated inundation. This is appropriate for event-scale emulation but does not resolve sub-daily rainfall intensity, antecedent hydrologic state, drainage-system dynamics, lead-time uncertainty, or sensor latency. Therefore, the present method should not yet be described as an operational nowcasting or early-warning system. Operational deployment would require sub-daily forcing, real-time data pipelines, sensor-failure handling, uncertainty calibration, out-of-distribution detection, and quantified inference latency.

\subsection*{Runtime benchmarking} The hydrodynamic simulation runtime reported in the introduction provides context for why a surrogate could be useful, but the present source package does not include an independently measured emulator inference-time benchmark. Before making strong speed-up claims, a final version should report the trained model size, hardware, batch size, preprocessing time, neural-network inference time, postprocessing time, and total wall-clock latency for one $256\times256$ event. This revision avoids inserting an unsupported numerical speed-up.

\subsection*{Outlook} Despite these limitations, the results suggest a useful direction for observation-guided simulation emulation: keep real observations out of the model input to avoid inference-time copying, but use them as targeted supervision to reduce local simulator bias. With multi-gauge experiments, independent flood-extent validation, ablation studies, calibrated uncertainty, and documented inference timing, this framework could become a practical component of rapid flood-risk assessment workflows.

\section*{Data and code availability}
The current preprint is accompanied by simulation-derived figures and this LaTeX source. The arXiv source package includes the manuscript source and figure files needed to compile the PDF. For a later peer-reviewed submission, the processed event list, train/test split definitions, USGS site and parameter identifiers, unit conversions, gauge preprocessing code, model-training scripts, loss-weight schedules, random seeds, inference-runtime benchmark, and plotting code should be released in a public repository or archived with a DOI, subject to any data-use restrictions for the underlying inputs.

\bibliographystyle{unsrt}
\bibliography{references_arxiv_final_fixed}

\end{document}